\documentclass[utf8]{frontiersSCNS}
\usepackage{url,lineno,microtype,subcaption}
\usepackage{soul}
\usepackage[onehalfspacing]{setspace}
\usepackage[hang, flushmargin]{footmisc}
\usepackage[colorlinks=true]{hyperref}
\usepackage{footnotebackref}

\def\keyFont{\fontsize{8}{11}\helveticabold }
\def\firstAuthorLast{Schubotz {et~al.}} 
\def\Authors{Moritz Schubotz\,$^{1,2,*}$, Ankit Satpute\,$^{1,2}$, Andr\'{e} Greiner-Petter\,$^{1}$, Akiko Aizawa\,$^{3}$ and Bela Gipp\,$^{1,4}$}
\def\Address{$^{1}$Chair for Data and Knowledge Engineering, University of Wuppertal, Wuppertal, Germany \\
$^{2}$ FIZ Karlsruhe - Leibniz Institute for Information Infrastructure, Berlin, Germany \\
$^{3}$Digital Content and Media Sciences Research Division, National Institute of Informatics, Tokyo, Japan \\
$^{4}$Chair for Scientific Information Analytics, University of Göttingen, Göttingen, Germany}

\def\corrAuthor{Moritz Schubotz}
\def\corrEmail{moritz.schubotz@fiz-karlsruhe.de}

\begin{document}
\onecolumn
\firstpage{1}

\title[Caching data science experiments]{Caching and Reproducibility: Making Data Science experiments faster and FAIRer} 

\author[\firstAuthorLast ]{\Authors} 
\address{} 
\correspondance{} 

\extraAuth{}

\maketitle

\begin{abstract}
Small to medium-scale data science experiments often rely on research software developed ad-hoc by individual scientists or small teams.
Often there is no time to make the research software fast, reusable, and open access.
The consequence is twofold. First, subsequent researchers must spend significant work hours building upon the proposed hypotheses or experimental framework.
In the worst case, others cannot reproduce the experiment and reuse the findings for subsequent research.
Second, suppose the ad-hoc research software fails during often long-running computational expensive experiments. In that case, the overall effort to iteratively improve the software and rerun the experiments creates significant time pressure on the researchers.

We suggest making caching an integral part of the research software development process, even before the first line of code is written.
This article outlines caching recommendations for developing research software in data science projects.
Our recommendations provide a perspective to circumvent common problems such as propriety dependence, speed, etc.
At the same time, caching contributes to the reproducibility of experiments in the open science workflow.
Concerning the four guiding principles, i.e., Findability, Accessibility, Interoperability, and Reusability (FAIR), we foresee that including the proposed recommendation in a research software development will make the data related to that software FAIRer for both machines and humans. 
We exhibit the usefulness of some of the proposed recommendations on our recently completed research software project in mathematical information retrieval.
\tiny
 \keyFont{ \section{Keywords:} Caching, Data Science, Reproducibility, Open Science, Research Software}
\end{abstract}

\section{Introduction}

A data science project involves exploratory data analysis, building hypotheses, performing experiments, and reporting results. 
Supportive intermediate steps of designing experimental framework, negative findings often remain unpublished due to the focus on central hypothesis proving. 
The open science perspective promotes publishing intermediate states or revisions and any unfavorable outcome.
Reproducibility is using published code and experiment-related resources to produce the reported results in a paper. 
\cite{Nosek2015} pointed out that reproducibility helps researchers gain different intuitions and build upon existing experimentation. 
Unpublished intermediate experiments and less focus on speeding up the experiments make the task of reproducibility difficult, and researchers have to spend significant work hours to prove the original contributions.
Top conferences in machine learning have introduced reproducibility challenges and voluntary code submission guidelines to improve reproducibility if the experiments were carried out to prove the hypothesis~\cite{Pineau2020}. 
However, evaluators observed that researchers faced difficulty reproducing the actual reported results in many cases.
Follow-up competitions (\text{ML Reproducibility Challenge 2021}\footnote{\url{https://paperswithcode.com/rc2021}}, \text{ML Reproducibility Challenge Spring-2021}\footnote{\url{https://wandb.ai/wandb_fc/reproducibility-challenge/reports/ML-Reproducibility-Challenge-Spring-2021}}) were organized by offering incentives to reproduce the results of the accepted papers.

A typical number of experiments involved in a project contributes to a load of reproducibility. 
A general caching approach helps the system speed the experiments by avoiding computationally expensive queries. 
Caching happens at multiple levels and is often considered a system designer's concept, ignored while developing software~\cite{Mertz2017}.
Advanced caching techniques exist such as \textbf{memoization}, i.e., an optimization technique used to enhance the speed of programs by storing the results of expensive functions and returning the cached result if the same inputs occur again~\cite{Mayfield1995}. 
However, researchers are more focused on the end goal, so adopting such techniques is ignored in developing research software.
Caching particular states of the application saves time and contributes to a system's independence from proprietary software or Application Programming Interfaces (APIs).
Researchers are focused on experimental results and ignore the procedures that could speed up their experiments.
There is also less motivation behind publishing the original experimental framework; hence applying performance improvement recommendations while developing software is often ignored.
To address common problems with reproducibility, \cite{Wilkinson2016} proposed FAIR principles.
The use of these principles is related to data and applies to algorithms, tools, experiment parameters, etc.
However, these recommendations in developing research software are yet to be explored.
In this work, we put our perspective of making experiments in a research software reproducible and faster, thus directly influencing FAIR principles.
We envision that the proposed recommendations will motivate researchers to conduct detailed analysis to find ways for making data related to research software FAIRer.

Caching has become highly crucial to applications, given its advantages in terms of speed. 
Systems nowadays consider cache memory size an integral part of architecture for applications running on top of it. 
On the system side, caching saves excess CPU usage at the cost of memory usage. 
On the user side, fast results could be presented instead of waiting to process the input. 
Furthermore, subsequent or same users can build upon these results to develop experiments if the experiments are permanently cached. 
Despite the popularity of caching, application-level caching involves designing routines to capture the application's specific states, and there are no general guidelines. 
This makes designing such routines difficult when developing software or experimental frameworks.
We draw out general recommendations from our experience with developing research software, which will help researchers build software capable of reproducing experiments.
We focus on making the recommendations not too detailed to focus on developing a specific type of software but to be easily able to adopt in the development cycle of a general case. 

\section{Background and Related Work}

In this work, we briefly mention the related caching approaches based on the developer's point of view.
Theoretically, one can cache every state of an experiment or data fed to the system. 
However, if the data to be cached is of high volume, it could quickly populate the memory, making it unavailable for new incoming data. 
Not every data coming in cached memory could stay forever because the memory is limited. 
We have cache eviction policies that decide which data to evict.
There are deterministic, randomized, optimization, and learning-based algorithms to decide upon eviction patterns~\cite{Jain2019}. 
Some \text{standard approaches}\footnote{\url{https://www.alibabacloud.com/blog/caching-essential-skills-for-developer_596213}} are available for caching when considered from a software developer's point of view. These are in-memory caching, web caching, CDN (Content Delivery Network) caching, and database caching. 

To the best of our knowledge, no previous study focuses on general caching practices used in data science projects. 
Also, there exists no standard set of experimental state caching guidelines for general cases. 
There are code wrappers for languages such as R (\text{Package: R.cache}\footnote{\url{https://github.com/HenrikBengtsson/R.cache}}), Python (\text{Python-memoization}\footnote{\url{https://github.com/lonelyenvoy/python-memoization}}), Java (\text{JCache}\footnote{\url{https://docs.oracle.com/en/middleware/standalone/coherence/14.1.1.0/develop-applications/introduction-coherence-jcache.html}}, \text{Caffeine}\footnote{\url{https://github.com/ben-manes/caffeine}}), which provides separate packages for caching, such as experimental states and outputs. However, extensibility of these packages in general research software development is not studied in any work yet.

\cite{Mertz2017} put forward a qualitative study of 10 web applications to see how developers handle caching logic in web applications.
They have shown that adaptive caching approaches possess tremendous potential. However, while designing software, it is not yet considered standard practice.
They issued a standard set of recommendations for developers of web-based projects; however, the authors did not comment on whether the recommendations were domain-specific or domain-neutral. 
\cite{Toffola2015} proposed an iterative dynamic analysis called \emph{memoizeIt} to discover methods that may benefit from memoization. 
The authors implemented the tool specifically for Java applications. 
Discussion of extension to other programming languages is not discussed.
What are the most observed patterns in commonly  developed software remains unanswered.

\section{Caching in data science}\label{sec:cachingdatascience}
In data science, repetitions of the same routines could be cached to speed the experiments or decrease data access latency.
The most straightforward adaptation policy is to take advantage of the locality of reference principle, i.e., the latest handled data is likely to be requested again. 
Caching in general applications involves four critical issues: (1) how to cache the selected data; (2) what data should be cached; (3) when the selected data should be cached or evicted; and (4) where the cached data should be placed and maintained. 
Even though application-level caching is being commonly adopted, there is no standard in determining how to approach these problems. Hence, most developers do it in an ad-hoc way~\cite{Mertz2017}. 
Algorithms in data science-related areas often rely on multiple state calculations, which takes time to produce an output. 
A typical data science project starts with raw data, and it goes through subsequent revisions such as cleaning, adjusting to a structure for experiments, etc.
\text{Data Version Control}\footnote{\url{https://dvc.org/doc}} could be an excellent match to track changes in data and allow anyone to access a specific revision, which would benefit other researchers to start performing new analyses at any intermediate state of the data.
One could conduct a study to find how to enrich the captured data revisions to make them easily findable and accessible.

\subsection{Caching Recommendations}

This section points out the general caching practices that one  should use while developing research software. We also mention observations coming from the development process of our software.
\newline
\textbf{Cache-First approach:} Based on our experience, it is much easier to consider caching while developing the software than doing in the final step of the development.
After requirement analysis, the user could combine ideas for caching particular states in the designing and prototyping step of software development. 
To be included with the cached data, one should decide upon the structure and constituents of metadata, such as the version of dependent software (at the time of caching), time to query, etc.
Adding metadata is particularly important if persistent caching is used, as it will help users to make the data accessible for further experiments.
It is also beneficial to assign global identifiers at the time of caching, contributing to the easy recognition of the data.
Cached data and additional metadata of the application's state could result in better search results, thus improving its Findability.
\newline
\textbf{Proprietary independence:} It is essential to cache interactions with proprietary software or API simply because the same researchers can use the cached results to build up their experiments.
If possible, then the smallest input (that could not be separated into individual components) to the propriety software should be identified and cached along with its output value.
Users should also perform persistent caching by permanently saving some states in the database, which would enhance interoperability of cached data.
A user could apply a similar principle in the case of expensive experiments both in terms of time and space. 
Decoupling software from such dependencies would make it easier to perform the local environment experiments, which would eventually benefit if the external software had undergone changes or revisions.
By caching such experimental states, we save efforts for future experiments for the current developer and the future reproducer.
\newline
\textbf{Web application:} If the software is of web application type, then the user should consider different hierarchies of caches. First, database caching practices should be used on the server-side to serve the frequent requests. Second, on the client-side, a user should use a browser-based cache. Third, the network cache should be maximized. 
Almost all the databases have some caching abilities in them. If the interaction with the database is frequent, then changing parameters of database-caching specific to the application might help reduce the query-serve time.
This is highly recommended, especially if the results from a query are returned as a file or more than just a number. 
Limited cache memory often produces challenges to cache a large quantity of data.
In the case of caching in cache-memory or local memory (should be faster than caching in a database), initially, one  should consider simple eviction policies such as the Least Recently Used (LRU).
If required, extra efforts should be put to deal with space optimization within cache memory, but based on our experience, choosing the appropriate framework requires a lot of time.
\newline 
\textbf{RESTful caching:} Software containing 
Representational state transfer (\text{REST}\footnote{\url{https://www.ics.uci.edu/~fielding/pubs/dissertation/rest_arch_style.htm}}) styles request handling could benefit by caching.
The architecture of REST allows users to cache data of a response to a request.
Users can implicitly or explicitly label data as cacheable or non-cacheable. 
If a response is cacheable, then a client cache is given the right to reuse that response data for later, equivalent requests.
The advantage of adding caching is that they completely or partially eliminate interactions, improving efficiency, scalability, and user-perceived performance~\cite{Fielding2000}.
However, caching comes at the cost of stale data if the cached data for the request differs significantly from the data present on the server for the same request.
\newline
\textbf{Big Data:} Experiments involving analytic engines such as Apache Spark (\href{https://scicrunch.org/resolver/RRID:SCR_016557}{RRID:SCR\_016557}) and Hadoop Distributed File System (\href{https://scicrunch.org/resolver/RRID:SCR_011879}{HDFS RRID:SCR\_011879}) should use caching provided by the sources.
Especially in \text{Spark}\footnote{\url{https://spark.apache.org/docs/3.0.0-preview/sql-ref-syntax-aux-cache.html}}, data-frames can be cached to perform repetitive operations without loading the data from HDFS or other storage interfaces such as Amazon S3 (\text{Simple Storage Service}\footnote{\url{https://aws.amazon.com/s3/}}).
There are size limitations, but it is recommended to use default caching in any case, and if the data exceeds the allowed cached memory, it is by default thrown into the primary storage.
\newline
\textbf{Black box caching:} Isolating interactions in the case of the black box part of software helps us get rid of uninfluenced behavior.
Caching in the black box procedures (in which internals are unknown) should be done for input-output pairs and their internal states if applicable.
On the user side, the time required to get the output from a black box should also be adjusted in data structure and used in subsequent runs; if the cached query is issued and if it has a large run time, then the user should be warned of the delays for such queries.
In the case of training neural networks or other models using Graphics Processing Unit (GPU) or Tensor Processing Unit (TPU), a local cache can also be used to save specific states, hyperparameters, or data. 
Almost all libraries that allows user to perform such experiments offer default caching.
For example, \text{Tensorflow}\footnote{\url{https://www.tensorflow.org/datasets/performances}} offers a small dataset (less than 1 GB) caching which helps reduce the time in processing within the multiple iterations. 
\newline
\textbf{Cache parameters monitoring:} Caching-related parameters such as cache-memory size, cache entry repetition, etc., should be monitored. 
Appropriate flags should be raised to the user if these parameters go to full capacity or are out of bounds.
In the case of persistent caches, the robustness of cached data should be checked if the dependent propriety software is changed or upgraded to a newer version.
We faced difficulties in using the persistent caches stored from the previous versions, especially when the new version's input format is changed; even though the entry for the requested data is present in the cache, it won't be used because of the format differences.

\subsection{Example case}\label{sec:eval}

This section demonstrates the use of some of the caching practices explained above and their advantages in developing a project.
In previous projects, we developed LaCASt, a \LaTeX{} to Computer Algebra System (CAS) translator~\cite{GreinerPetter2019}.
LaCASt has the option to interact with CAS such as Maple or Mathematica in order to evaluate a translated expression~\cite{Cohl2018,Greiner-Petter2022}.
CASs are mathematical software tools with a wide variety of applications, from numeric evaluations to symbolic transformations.
Interactions of LaCASt with the CAS can be isolated from the main translator and could be cached to be reused for further experiments.
To decide the data structure of the interactions to be cached, we identified the nature of queries to CAS from LaCASt.
In the following, we will elaborate on our caching strategy for LaCASt as an example use case for the outlined caching strategies from above.
Note that we will only mention the steps of LaCASt's workflow that are related to caching. 
A detailed evaluation pipeline is provided in~\cite{Cohl2018,Greiner-Petter2022}.

LaCASt uses independent \emph{evaluation cycles} to evaluate a translated formula.
In each evaluation cycle, LaCASt constantly interacts with the CAS.
These interactions are slow because CASs are heavy and large software packages.
In addition, each input to a CAS may have an impact on following interactions.
Therefore, caching an interaction with a CAS would require caching all previous inputs to the CAS too.
However, an evaluation cycle used by LaCASt has no effect on other evaluation cycles because all inputs to the CAS are reset at the end of a cycle, i.e., they are independent.
This independence of the evaluation cycles allows us to implement effective caching.
Cached data along with additional information about the state and nature of the evaluation cycle would make it discoverable for further experiments.

\begin{flushleft}
\captionof{table}{LaCASt's implemented characteristics that influences FAIR principles.}\label{table:relatednessFAIR} 
\begin{tabular}{|p{0.19\linewidth}|p{0.73\linewidth}| } 
\hline
Guiding principle & Implementation in LaCASt \\
\hline
    F – Findability & LaCASt’s cache contains all formulae in LaTeX, Maple and Mathematica code including additional meta information, such as applied constraints. By publishing this cache, other scientists potentially find LaCASt by searching for famous mathematical formulae or their Computer Algebra System representations.\\
\hline
    A – Accessibility & Thanks to caching, we decoupled LaCASt from the proprietary API of Maple and Mathematica. This, in turn, allows for using LaCASt without the CAS running in the background. In turn, most functionality of LaCASt is accessible without the need of proprietary software.\\
\hline
    I – Interoperability & The decoupling of LaCASt from the CAS Maple and Mathematica also allows to replace the proprietary solutions with open-source alternatives, such as the CAS SymPy (\cite{SymPy}). Allow replacing the CAS with other solutions greatly improves the interoperability of LaCASt.\\
\hline
    R – Reproducibility & The cached computations allow other scientists to exactly re-run our experiments even if Maple or Mathematica are unavailable. Hence, we enabled the reproducibility of LaCASt on our dataset. In addition, the cache also persists the current state of the CAS on our test dataset. Hence, the cache allows us to analyze improvements in newer CAS versions or even compare the performance with entirely different CAS, such as the aforementioned open-source alternative SymPy.\\
\hline
\end{tabular}
\end{flushleft}

We cache the CAS inputs and outputs of each step within each evaluation cycle.
The cached entries (the CAS inputs and outputs) of a single step only depend on the cached entries of the previous steps within the same cycle.
Our cache data structure is a simple key-value map (keys are the inputs to the CAS and the values are the outputs to this input) in which each key recursively depends on the previous keys within the same evaluation cycle.
The independence of the evaluation cycles also allows us to add an additional useful property to our cache system:
two inputs $i_k, j_k$ in two distinct evaluation cycles $C_1, C_2$ will return the same value if and only if all previous inputs in the two evaluation cycles were identical, i.e., $i_l = j_l \ \forall \ l < k$ with $i_l \in C_1$ and $j_l \in C_2$.
This property significantly reduces the memory footprint of our cache because we can store key-value pairs across multiple evaluation cycles as long as the previous steps within these cycles were identical.
In the future, one can verify a new formula using already cached (persistent) interactions.
If the interactions of the new formula are present in the cache, then it is a cache hit else cache miss.
One could measure the performance gain using the caching recommendations in the LaCASt quantitatively.
For example, a similar experiment as shown in \cite{Padulano2021} could be conducted with LaCASt.
The authors used caching techniques to show that the average speedup of factor 2 was obtained in distributed RDataFrame after caching was enabled.
This article does not evaluate LaCASt's performance gain using caching, but we direct a way to make experiments in LaCASt more FAIRer.

By caching the states, we achieved the following goals.
First, we eliminated the dependency on costly proprietary software (most CAS charge users after a trial period).
Second, using persistent cached experiments, users can perform verification of new translated expressions at any point in the future.
Third, directly available results save a user time in performing experiments.
Fourth, if the version of CAS is upgraded, we no longer have to develop new interaction types since our cached results are persistent.
Last but not least, the overall gains, when coupled with indexed metadata, would contribute to making the data related to the application more FAIRer.
Table \ref{table:relatednessFAIR} illustrates in brief, how caching in LaCASt along with other characteristics contributes towards the FAIRness of the research software.

\section{Conclusion and Future Work}

We outlined the caching practices for the general data science project, which helps make experiments in research software reproducible and puts a step toward making experiments FAIRer .
The proposed broad set of guidelines will give developers a starting point to include simple caching practices while developing a research software or experimental setup for a hypothesis proving.
Using an example software development process of our research project, we showed that some caching practices improve the reproducibility of results and could make experiments FAIRer by extension.
Analysis of outlined caching practices based on parameters such as scalability and availability will be a promising aspect to know the practical limitations in use cases involving heavy load.
In the future, it would be beneficial to thoroughly survey some research software used in data science projects to see how many of them make use of proposed caching recommendations.
This process will help us improve the recommendations if the parts of the studied software can be cached and categorized in any other caching practices.
In this way, we will have a standard set of categories based on the practical observations from projects developed by other researchers.
The proposed recommendations contribute directly to reproducibility and briefly towards FAIR aspects.
However, in the future work, we would like to evaluate if and how existing software uses FAIR guidelines.
Follow-up work would involve proposing recommendations to make FAIR guidelines easy to adopt in research software.
We observed that conferences in machine learning have provided reproducibility tasks for the submissions and pointed out obstacles in reproducing the results. 
However, it would be essential to point out how many of the published approaches uses caching to help the reproducibility of results.
Recommendations from the reproducibility challenge and outlined caching practices from this work will evaluate the impact on the quality of the research.

\section*{Conflict of Interest Statement}
The authors declare that the research was conducted in the absence of any commercial or financial relationships that could be construed as a potential conflict of interest.

\section*{Funding}
Our research was supported by the German Research Foundation (DFG grant GI 1259/1) and the IFI program of the German Academic Exchange Service (DAAD grant no.: 57515245).

\section*{Acknowledgments}
We thank J\"{u}rgen Gerhard from Maplesoft for his continuous support.

\bibliographystyle{frontiersinSCNS_ENG_HUMS}

\bibliography{test}

\end{document}